\documentclass{article}
\usepackage{amsmath,amssymb,amsthm,spconf,xcolor}
\usepackage[pdftex]{graphicx}
\usepackage{fixmath}

\newcommand{\vol}{\operatorname{vol}}
\newcommand{\tr}{\operatorname{Tr}}
\newcommand{\sE}{\mathsf{E}}
\newcommand{\bbR}{\mathbb{R}}
\newcommand{\bbI}{\mathbb{I}}

\newcommand{\tx}{\tilde{x}}
\newcommand{\ty}{\tilde{y}}

\newcommand{\cN}{\mathcal{N}}
\newcommand{\cM}{\mathcal{M}}

\newcommand{\cF}{\mathcal{F}}
\newcommand{\DKL}{\Delta_{\mbox{\tiny KL}}}
\newcommand{\DMI}{\Delta_{\mbox{\tiny MI}}}
\newcommand{\pd}[1]{\frac{\partial}{\partial #1 }}

\title{An Information-Geometric Approach to Sensor Management}
\name{B.~Moran${}^1$, S.~D.~Howard${\,}^2$ and D.~Cochran${}^3$}
\address{${}^1$University of Melbourne, Parkville VIC, Australia\\${}^2$Defence Science and Technology Organisation, Edinburgh SA, Australia\\${}^3$Arizona State University, Tempe AZ, USA}
\begin{document}
\ninept
\maketitle
\begin{abstract}
An information-geometric approach to sensor management is introduced that is based on following geodesic curves in a manifold of possible sensor configurations.  This perspective arises by observing that, given a parameter estimation problem to be addressed through management of sensor assets, any particular sensor configuration corresponds to a Riemannian metric on the parameter manifold. With this perspective, managing sensors involves navigation on the space of all Riemannian metrics on the parameter manifold, which is itself a Riemannian manifold. Existing work assumes the metric on the parameter manifold is one that, in statistical terms, corresponds to a Jeffreys prior on the parameter to be estimated. It is observed that informative priors, as arise in sensor management, can also be accommodated. Given an initial sensor configuration, the trajectory along which to move in sensor configuration space to gather most information is seen to be locally defined by the geodesic structure of this manifold. Further, divergences based on Fisher and Shannon information lead to the same Riemannian metric and geodesics.
\end{abstract}

\section{Introduction}
\label{sec:intro}

The work of Amari and others \cite{Amari2000} on the use of methods of Riemannian geometry to analyze statistical estimation problems is of increasing interest to researchers in signal processing.  This methodology, known as {\em information geometry}, provides a rigorous framework for measuring the power of data to discriminate values of parameters. These ideas date back to Rao \cite{Rao45}, who showed that the Fisher information of a likelihood used in an estimation problem can be seen as a Riemannian metric on the parameter manifold. 

This paper brings an information-geometric perspective to a class of sensor management problems by casting the objective of sensor management as parameter estimation and describing how this leads to the role of sensor management as selecting a Riemannian metric for the parameter manifold.  Established results in Riemannian geometry \cite{gil-medrano91}, outside the context of information geometry, show that the collection of all Riemannian metrics on a Riemannian manifold is itself an (infinite-dimensional) Riemannian manifold.  In problems where the collection of possible sensor actions is suitably modeled by a smooth finite-dimensional manifold, the space of interest is a finite-dimensional sub-manifold of this infinite-dimensional Riemannian manifold. A perspective is developed in which the best sensor management action to take, in terms of gathering the most information relevant to the estimation objective, is locally characterised in terms of geodesic curves in this space.

Much of the development in subsequent sections of this paper is rather abstract and draws upon mathematical machinery that is unfamiliar to many researchers in sensor management area.  To provide a more concrete context in which to illustrate some of the concepts that arise in later sections, is it helpful to begin by setting forth an example problem.  Suppose two mobile sensor platforms and one stationary target (emitter) are located in the plane $\bbR^2$, as depicted in Fig.~1. The goal is to estimate the position of the target from bearings-only measurements taken at the sensors.  Since the sensors are mobile, the sensor management problem is to identify the trajectories of sensor motion that will yield the best estimate of the target position.
\begin{figure}[hbt]
\begin{center}
\includegraphics[width=0.65\linewidth]{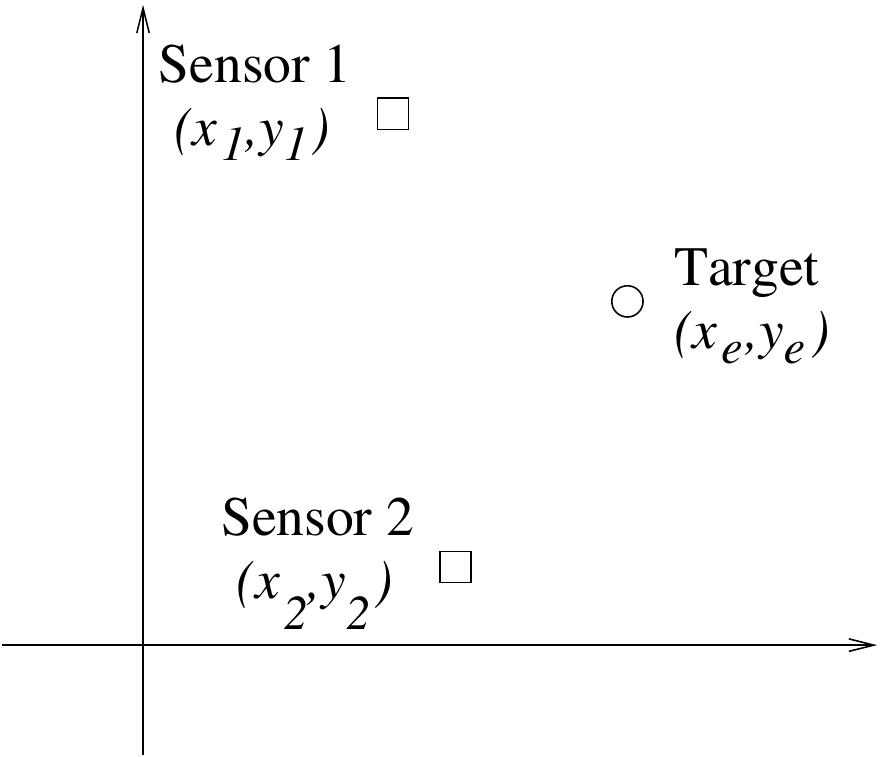}
\end{center}
\caption{An illustrative scenario involves estimating the position $(x_e,y_e)$ of a single stationary emitter from bearings-only measurements received at two mobile sensing platforms located at respective coordinates $(x_1,y_1)$ and $(x_2,y_2)$ in the plane.}
\end{figure}
More specifically, the target position is $(x_e,y_e)$ and the sensor positions are $(x_j,y_j)$ for $j=1,2$.  Denoting $\tx_j=x_j-x_e$ and $\ty_j=y_j-y_e$, the bearing of the target from sensor $j$ is $\varphi_j=\arctan(\ty_j/\tx_j)$. The sensor measurements are independent and von Mises distributed, each with common inverse dispersion parameter $\kappa$ and with the circular mean of the measurement at sensor $j$ having circular mean $\varphi_j$. 

The following sections proceed to describe the nature of a sensor model from an information-geometric viewpoint, to define the parameter manifold, the sensor manifold, its metric structure, and to derive a differential equation that characterizes geodesic curves on the sensor manifold. The development departs from the purely geometric treatment in \cite{gil-medrano91} in that it allows for informative prior distributions on the parameter manifold rather than restricting attention to a volume form corresponding to the Jeffreys prior.  Further, the Riemannian metric with respect to which geodesics maximize ``energy integrals'' on the sensor manifold is shown to arise from both Kullback-Leibler and mutual information perspectives.  Throughout this process, the example just introduced will be used to illustrate these concepts in a concrete fashion.

\section{Sensor Model}
\label{sec:model}

Consider the problem of estimating a parameter $\theta$ from data $x$ collected by sensors up to time $t$. Beginning with a prior probability distribution for $\theta$, which may reflect what is known from previous measurements or side information, the effect of taking measurement at time $t$ is to provide a posterior probability distribution, which will be assumed to be represented by a posterior probability density $p(\theta|x)$.  If the option exists to use one of a parametrized set of sensors or sensor configurations, each of these will produce its own posterior density.  When these posteriors are known, selecting a sensor configuration to use for a measurement amounts to choosing one of them from which to estimate $\theta$. The parameter $\theta$ will henceforth be assumed to be an element of an $m$-dimensional smooth manifold $M$ ($C^\infty$ will be assumed, although $C^2$ is sufficient for most of the discussion here), which will be called the ``parameter manifold.''  In the example problem, $M$ is $\bbR^2$ because the parameter $\theta=(x_e, y_e)$ is a physical location in the $x-y$ plane.

Denoting by $\ell =\log p(x|\theta)$ the log-likelihood for this problem, the Fisher information is
\begin{equation}
  \label{eq:fisher_information}
  \cF_\theta = \sE_{p(\cdot|\theta)}[d_\theta\ell\otimes d_\theta\ell] ,
\end{equation}
where $d_\theta\ell$ denotes the derivative of $\ell$ with respect to the parameter $\theta$. This is well known to be equivalently expressed as
\begin{equation*}
   \cF_\theta = -\sE_{p(\cdot|\theta)}[\nabla^2_\theta\ell] .
\end{equation*}
In this expression, $\nabla_\theta$ represents the covariant derivative along any connection in $M$; all choices of connection give the same quantity as \eqref{eq:fisher_information}.  
$F_\theta$ is always a non-negative definite $m\times m$ matrix, and in what follows it will be assumed to be non-singular, thus giving rise to a continuous family of inner products over the tangent spaces of the manifold. Direct calculation in the example problem shows that, in the coordinate system described above and depicted in Fig.~1, $F_\theta$ has the form
\begin{align}
\label{eq:example_fisher}
  \cF_\theta & =\kappa A(\kappa)\sum_{j=1}^2 \frac{1}{R_j^4}
  \begin{pmatrix}
    \ty_j^2 & -\tx_j \ty_j \\ \nonumber
    -\tx_j\ty_j & \tx_j^2
  \end{pmatrix}\\
 &= \kappa A(\kappa)\sum_j\frac{1}{R_j^4}(\ty_j,-\tx_j)\otimes(\ty_j,-\tx_j)\\ 
 &= \kappa A(\kappa)\sum_j\frac{1}{R_j^2}(\sin \varphi_j,-\cos\varphi_j)
    \otimes (\sin\varphi_j, -\cos \varphi_j) , \nonumber
\end{align}
where $R_j^2=\tx_j^2+\ty_j^2$.

Through this mechanism, the choice of a particular sensor leads to the association of a positive definite matrix with each $\theta\in M$, thereby imbuing $M$ with a \emph{Riemannian metric} that measures the ability of that sensor's data, at least locally, to discriminate between parameter values. It is possible to calculate the shortest distance, in terms of this metric, between two values $\theta$. As discussed in \cite{Amari2000}, the Kullback-Leibler divergence between $p(x|\theta)$ and $p(x|\theta')$ is approximately half of the square of the distance between $\theta$ and
$\theta'$ when $\theta$ and $\theta'$ are close.

\section{The Sensor Manifold}

It has been shown \cite{gil-medrano91} that the collection $\cM(M)$ of all Riemannian metrics on the manifold $M$ is an infinite-dimensional (weak) Riemannian manifold. The structure of its tangent space is described in \cite{gil-medrano91}.  A point in $\cM$ is a Riemannian metric on $M$; i.e., it associates a positive definite form $g_\theta$ with each $\theta\in M$. Under suitable assumptions, a metric on $\cM$ is defined by 
\begin{equation}
  \label{eq:gil_metric}
  G_g(h,k)=\int_M \tr(g_\theta^{-1}h_\theta g_\theta^{-1}k_\theta)\; \vol(g_\theta) ,
\end{equation}
where $\vol(g_\theta)=\sqrt{\det(g_\theta)}\,d\theta$. Specific assumptions guaranteeing finiteness of this integral are beyond the scope of this discussion, and it will suffice for the purposes here to assume directly that it is finite. 

Although the nature of $\cM$ appears formidable, realistic sensor management problems do not require one to work with this entire space, rather with a finite-dimensional sub-manifold that inherits the metric \eqref{eq:gil_metric} from $\cM$. The assumption that leads to this situation is that the collection of all possible sensor configurations is parametrized by a smooth manifold $S$, which will be called the ``sensor manifold.''  In the example problem, the sensor configuration is completely specified by the positions of the two sensor platforms in the plane; i.e., by $\sigma=(x_1,y_1,x_2,y_2)\in\bbR^4$. In this case, the sensor manifold is $S=\bbR^4$ and the only elements of $\cM$ of relevance are those metrics on $M$ that arise from a sensor configuration $\sigma$ in this four-dimensional manifold.

Beginning with a sensor configuration $\sigma\in S$ gives rise first to a likelihood $p_\sigma(x|\theta)$ and consequently to a Riemannian metric $g(\sigma)$ on the parameter manifold $M$, as described in Section \ref{sec:model}.  As a Riemannian metric on $M$, $g(\sigma)$ is an element of $\cM(M)$.  This mapping $g:S\to\cM$ taking $\sigma$ to $g(\sigma)$ will be called the ``sensor geometry.''  In what follows, $g$ will be assumed to be smooth and one-to-one and $g(S)$ a sub-manifold of $\cM$. Weaker assumptions are possible (e.g., $g$ is an immersion), but full generality is is not needed here to adequately illustrate the method. Through the sensor geometry map, the finite-dimensional manifold $S$ inherits the Riemannian structure of $\cM$; i.e., the distance between two sensor configurations in $\sigma_1$ and $\sigma_2$ in $S$ is taken to be the distance between $g(\sigma_1)$ and $g(\sigma_2)$ in $\cM$. This construction endows the sensor manifold $S$ with its own Riemannian metric which captures, in information-theoretic terms, the ``complementariness'' of sensor configurations. 

\begin{figure}[hbt]
\begin{center}
\includegraphics[width=0.85\linewidth]{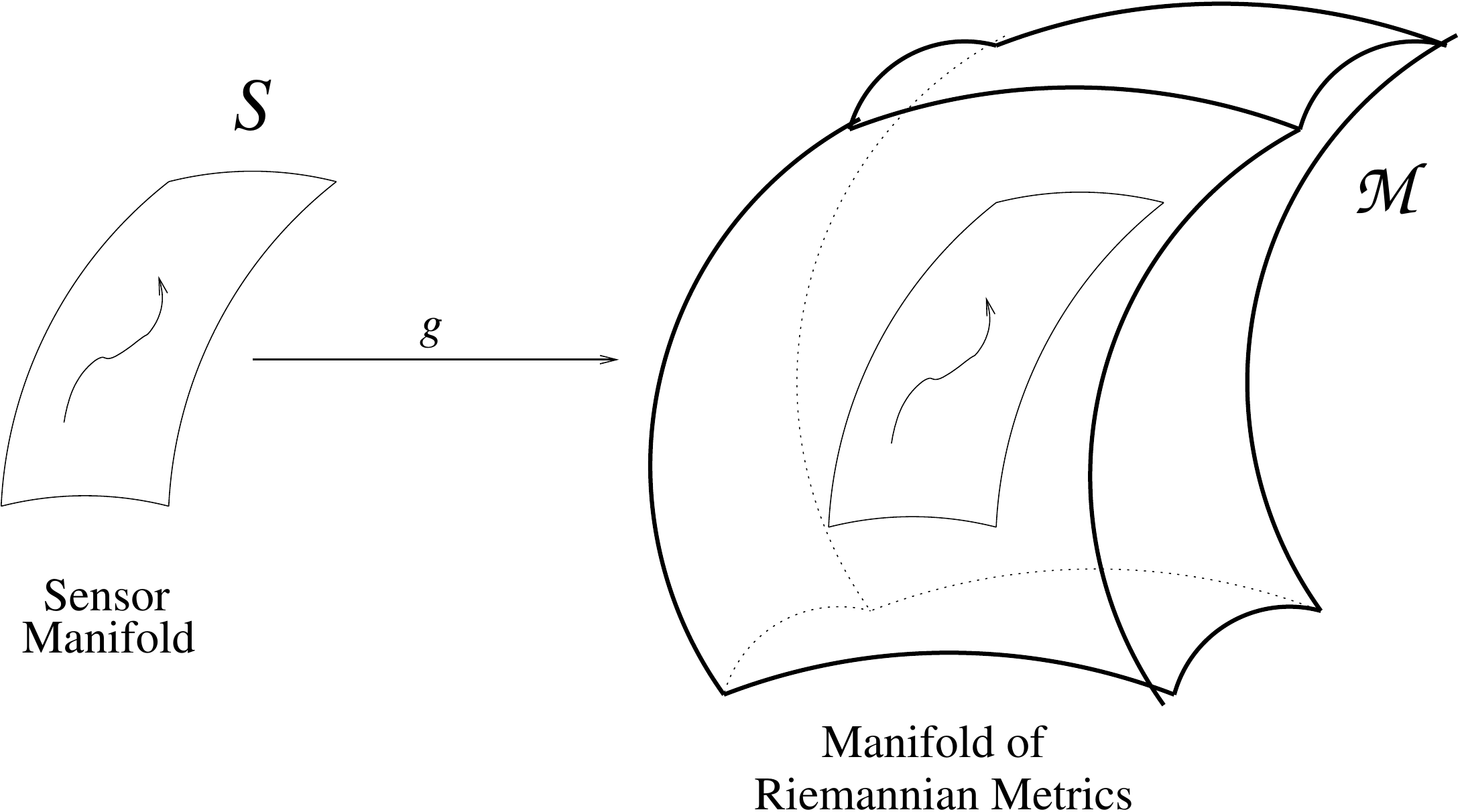}
\end{center}
\caption{The sensor geometry map $g$ allows the finite-dimension manifold $S$ of sensor configurations to inherit a Riemannian metric from the infinite-dimensional manifold $\cM$ of all Riemannian metrics on the parameter manifold $M$.}
\end{figure}

\section{Geodesics}
\label{sec:geodesics}

The objective of determining good trajectories in the sensor manifold $S$ will be addressed by relating these to geodesic curves in $\cM$. Following \cite{gil-medrano91}, consider a smooth curve $\gamma:[0,1]\to\cM$.  For each $t\in [0,1]$, $\gamma(t)$ is a Riemannian metric on the parameter manifold $M$ and thus associates a positive definite matrix $\gamma(t)_\theta$ with each point in $\theta\in M$. The energy integral along the curve $\gamma$ is
\begin{equation}
  \label{eq:energy_int}
E_\gamma=\frac{1}{2}\int_0^1\int_M \tr\bigl(\gamma^{-1}\dot{\gamma}
  \gamma^{-1}\dot{\gamma})\bigr)\;dF(\theta)\,dt .
\end{equation}
In this expression, $dF(\cdot)$ is a probability density on $M$, $\gamma$ means $\gamma(t)_\theta$, and $\dot{\gamma}$ is the derivative of $\gamma$ with respect to $t$. Geodesics in $\cM$ minimize $E_\gamma$, and a variational approach is used in \cite{gil-medrano91} to obtain the differential equation $\ddot{\gamma}=\dot{\gamma}\gamma^{-1}\dot{\gamma}$ for $\gamma(t)$, which implies
\begin{equation*}
\gamma(t)=\gamma(0)\exp(\gamma(0)^{-1}\dot{\gamma}(0)t)
\end{equation*}
The right-hand side of this differential equation is observed to be a Christoffel symbol.


The induced metric at a point $\sigma\in S$ is
\begin{equation*}
G_\sigma(u,v)=\int_M \tr\left(g(\sigma)^{-1}g_*(u)g(\sigma)^{-1}g_*(v)      \right)\,dF(\theta)
\end{equation*}
where $u$ and $v$ are in the tangent space $TS_\sigma$ of $S$ at $\sigma$ and $g_*$ is the push-forward of $g:S\rightarrow \cM$.  For a smooth curve $\gamma:[0,1]\rightarrow S$, the energy integral restricts to
\begin{equation*}
E_\gamma 
= \frac{1}{2} \int_0^1 \int_M \tr\left(g(\gamma(t))^{-1} g_*(\dot\gamma(t)) g(\gamma(t))^{-1} g_*(\dot\gamma(t))\right) dF(\theta) dt
\end{equation*}
The geodesics, which are the extremal curves of $E_\gamma$, satisfy
\begin{equation*}	
\ddot\gamma = -\Gamma_\gamma(\dot\gamma, \dot\gamma)
\end{equation*}
where $\Gamma$ denotes the Christoffel symbol for the Levi-Civita connection on $S$. 

In terms of local coordinates in $S$, geodesic equation in $S$ may be obtained by solving a variational problem on the path $u$.  To set this up, it is convenient to abuse notation and define a smooth function $u:[0,1]^2\to S$ such that $u(s,t)|_{s=0}=u(t)$.  With this notation, and coordinatizing $G_\sigma$ as $Q_{i,j}$,
\begin{multline*}
\left. \frac{\partial}{\partial s}\right|_{s=0} E_g
 = \frac{1}{2}\int_0^1 \left. \frac{\partial}{\partial s}\right|_0 \sum_{i,j}Q_{i,j}(u)u_t^i u_t^j\;dt \\
  = \frac{1}{2}\int_0^1 \left( \sum_{i,j,k} \partial_k Q_{i,j}(u) u_s^k u_t^i u_t^j \right. \\
  + \left. 2\sum_{i,j}Q_{i,j}(u)u_{ts}^i u_t^j \right)\;dt
\end{multline*}
If this expression is set to zero, further algebraic simplification leads to a differential equation (in coordinates) that characterizes geodesics in $S$:
\begin{equation*}
u_{tt}^\ell = \sum_{i,j}\left( -\sum_k Q^{\ell,k}\partial_i Q_{k,j}
+ \frac{1}{2}\sum_k Q^{\ell,k}\partial_k Q_{i,j} \right) u_t^i u_t^j .
\end{equation*}

Returning to the example pictured in Fig.~1, the local coordinates in $S=\bbR^4$ are $x_1$, $y_1$, $x_2$, and $y_2$. The positive definite matrix $g(u)$ corresponds to the Fisher information matrix $\cF_\theta$ given in \eqref{eq:example_fisher}.  The inverses and derivatives needed are calculable, with
\begin{multline*}
  \cF_\theta^{-1}=\left(\frac{R_1R_2}{\kappa
    A(\kappa)\sin^2(\varphi_1-\varphi_2)}\right) \times \\
    \sum_j \frac{1}{R_j^2}
  \begin{pmatrix}
    -\sin^2\varphi_j& \sin \varphi_j\cos \varphi_j\\
     \sin \varphi_j\cos \varphi_j& -\cos^2\varphi_j
   \end{pmatrix}
\end{multline*}
and
\begin{equation*}
  \cF'_\theta=\sum_j \pd{x_j}\cF_\theta \dot{x_j}+\sum_j\pd{y_j}\cF_\theta \dot{y}_j ,
\end{equation*}
where
\begin{align*}
  \pd{x_j}\cF_\theta &=\frac{\kappa A(\kappa)}{R_j^3}
  \begin{pmatrix}
    0&\sin \varphi_j\\
    \sin \varphi_j&-2\cos\varphi_j
  \end{pmatrix}\\
  \pd{y_j}\cF_\theta &=\frac{\kappa A(\kappa)}{R_j^3}
  \begin{pmatrix}
    -2\sin \varphi_j&\cos \varphi_j\\
    \cos \varphi_j&0
  \end{pmatrix}\\
\end{align*}

\begin{figure}[hbt]
\label{fig:example}
\begin{center}
\includegraphics[width=1.0\linewidth]{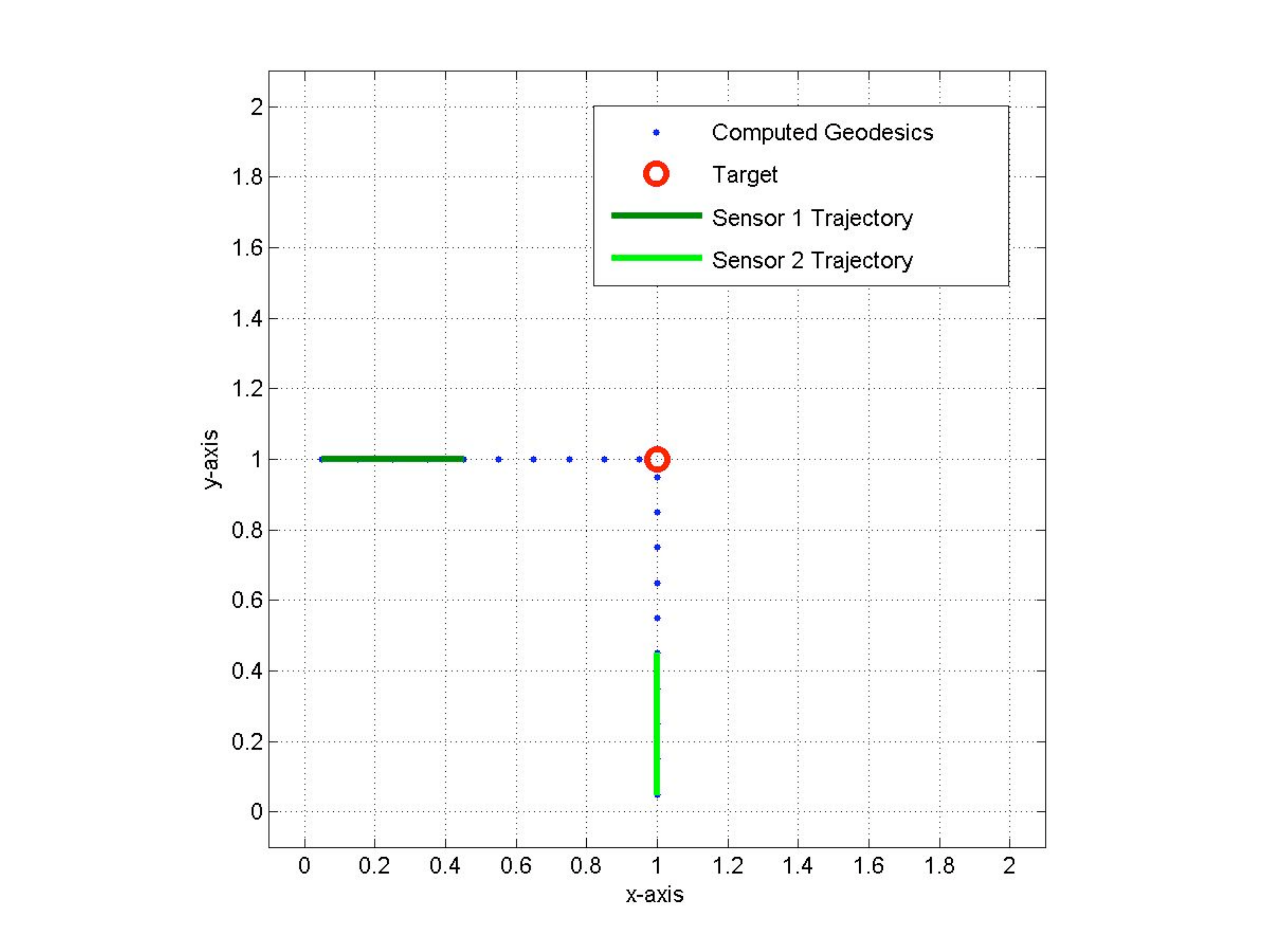}
\end{center}
\caption{Sensor trajectories based on geodesic approximation for the example scenario.  Sensor 1 starts at (0,1) and Sensor 2 starts at (1,0).  The target is at (1,1).}
\end{figure}

Fig.~\ref{fig:example} shows trajectories obtained for a particular case of the example scenario. The target is stationary at (1,1), and the sensors' prior distribution on the target location is normal with mean (1,1) and covariance $0.01\bbI$.  Sensor 1 starts at (0,1) and Sensor 2 starts at (1,0), and initial directions of motion are defined by the geodesic for this configuration.  The sensors move in this direction for a fixed period of time, a new set of directions is determined from geodesic calculations based on the new configuration, the sensors move again, and so forth.  The dotted trajectories are extrapolations; they indicate the directions defined by the geodesic computation at the last iteration computed.

\section{Divergences on $\cM$}
\label{sec:divergences}

The proposed scheme for sensor management involves following, at least locally, geodesic curves in $S$ defined by the Riemannian metric $S$ inherits from $\cM$.  Geodesics maximize energy integrals of the form \eqref{eq:energy_int}, so it is desirable to understand how optimization in this sense relates to the amount of information gathered by the sensor.  Consider first the Kullback-Leibler divergence $D(\cN_2 || \cN_1)$ for two multivariate normal distributions with equal means and respective non-singular covariance matrices $g$ and $h$.  This is given by 
\[
\frac{1}{2}\tr(g h^{-1}-\bbI)+\frac{1}{2}\log \frac{|h|}{|g|}
\]
where $|\cdot|$ denotes determinant and $\bbI$ is the identity matrix. A divergence on $\cM$ may be defined by
\begin{multline*}
\DKL(g,h) \\
= \int_M \left[\frac{1}{2} \tr\left(gh^{-1}-\bbI\right) +\frac{1}{2}\log\left(\frac{|h|}{|g|}\right)\right] \;dF(\theta) .
\end{multline*}
Here, the two positive definite matrices $g$ and $h$ are regarded as arising at each point of $M$ from two Riemannian metrics.  It is evident that $\DKL(g,g)=0$ and $\partial_g \DKL(g,h)|_{h=g} = \partial_h \DKL(g,h)|_{h=g} = 0$.  The corresponding Riemannian metric on $\cM$ is 
\[
\partial^2_g \DKL |_{h=g} = \partial^2_h \DKL |_{h=g} = \frac{1}{2}\int_M \tr(g^{-1} g' g^{-1} g') \;\; dF(\theta) ,
\]
as appears in \eqref{eq:energy_int}.
 
Similarly, one can define a divergence on $\cM$ motivated by mutual information by
\begin{multline*}
\DMI(g,h) = \int_M \left\{\log \left( \left| \frac{1}{2} (\bbI+g^{-1}h)\right| \right) \right. \\
+ \left. \log \left( \left| \frac{1}{2} (\bbI+h^{-1}g)\right| \right) \right\}\;dF(\theta) 
\end{multline*}
This ``symmetrized'' mutual information expression is equivalent to
\begin{multline*}
\DMI(g,h) = \int_M \left\{\log \left( \left| \frac{1}{2} (\bbI+g^{-1}h)\right| \right) \right. \\
+ \left.  \frac{1}{2} \log \left(\frac{|g|}{|h|}\right) \right\} \;dF(\theta)
\end{multline*}
As with $\DKL$, it is clear that $\DMI(g,g)=0$. Calculation reveals that $\partial_g \DMI(g,h)|_{h=g} = \partial_h \DMI(g,h)|_{h=g} = 0$ and that the corresponding Riemannian metric on $\cM$ is 
\[
\partial^2_g \DMI |_{h=g} = \partial^2_h \DMI |_{h=g} = \frac{1}{2}\int_M \tr(g^{-1} g' g^{-1} g') \;\; dF(\theta)
\]
Thus, despite arising from different concepts of information (i.e., $\DKL$ from Fisher and $\DMI$ from Shannon), both of these divergences give rise to exactly the Riemannian metric on $\cM$ used in the geodesic computations of Section \ref{sec:geodesics}.

\section{Conclusion}

In this short paper, we have built upon results in differential geometry, outside the context of information geometry, to introduce an information-geometric approach to sensor management. The approach begins with the observation that, when the goal of sensing is parameter estimation, the effect of selecting a particular sensor configuration amounts to imparting a Riemannian metric on the parameter manifold $M$ via the Fisher information.  The collection of all such metrics is the Riemannian manifold $\cM(M)$, for which the metric, geodesic equations, and other differential geometric aspects are known. With the assumption that our choices of sensor configuration are parametrized by a smooth ``sensor manifold" $S$, we observed that $S$ inherits a Riemannian structure from $\cM$ and used this to obtain a differential equation characterizing geodesic curves in $S$.  In  the purely geometrical work on which we have built, the measure on $M$ is a volume form that corresponds to the statistical notion of a (minimally informative) Jeffreys prior.  We observe that this may be replaced by an informative prior, as would typically be desirable in sensor management applications. Navigation along geodesic curves in a Riemannian manifold maximizes an energy integral involving the metric.  We have constructed two distinct divergences on $\cM$ corresponding to familiar information-theoretic quantities (Kullback-Leibler divergence and mutual information) that have been used by various authors as criteria in designing sensor scheduling algorithms.  Both of these are shown to lead to the same Riemannian metric on $\cM$, suggesting the information gathering merit of sensor scheduling based on following geodesic curves defined with respect to this metric.

While the work presented here is mostly conceptual, we have shown enough specifics of how the proposed method manifests in a concrete example to indicate its feasibility.  We are continuing to develop complete application examples while simultaneously working out rigorous specifics of some of the mathematical foundations.

\section{Acknowledgments}

The authors are grateful to Sofia Suvorova who supplied the numerical results presented in Sec.~\ref{sec:geodesics} in response to reviewer remarks on our original manuscript.  We regret that ICASSP policy prevents us from including her as an author on this revised version of the paper.

This work was supported in part by the University of Michigan and the U.S. Army Research Office under MURI award No.~W911NF-11-1-0391 and by the U.S.~Air Force Office of Scientific research under Grant No.~FA9550-09-1-0561.


\end{document}